# An NBDMMM Algorithm Based Framework for Allocation of Resources in Cloud

Mansaf Alam and Kashish Ara Shakil

**Abstract**—Cloud computing is a technological advancement in the arena of computing and has taken the utility vision of computing a step further by providing computing resources such as network, storage, compute capacity and servers, as a service via an internet connection. These services are provided to the users in a pay per use manner subjected to the amount of usage of these resources by the cloud users. Since the usage of these resources is done in an elastic manner thus an on demand provisioning of these resources is the driving force behind the entire cloud computing infrastructure therefore the maintenance of these resources is a decisive task that must be taken into account. Eventually, infrastructure level performance monitoring and enhancement is also important. This paper proposes a framework for allocation of resources in a cloud based environment thereby leading to an infrastructure level enhancement of performance in a cloud environment. The framework is divided into four stages Stage 1: Cloud service provider monitors the infrastructure level pattern of usage of resources and behavior of the cloud users. Stage 2: Report the monitoring activities about the usage to cloud service providers. Stage 3: Apply proposed Network Bandwidth Dependent DMMM algorithm .Stage 4: Allocate resources or provide services to cloud users, thereby leading to infrastructure level performance enhancement and efficient management of resources. Analysis of resource usage pattern is considered as an important factor for proper allocation of resources by the service providers, in this paper Google cluster trace has been used for accessing the resource usage pattern in cloud. Experiments have been conducted on cloudsim simulation framework and the results reveal that NBDMMM algorithm improvises allocation of resources in a virtualized cloud.

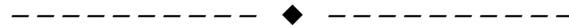

## 1 INTRODUCTION

CLOUD computing is the latest development in the field of computing. It has revolutionized a change in the way in which computing resources are being used. It is a fine blend of utility, virtualization and grid computing along with several other additional features such as providing computing resources as services, along with dynamic provisioning of resources and location independent access to resources. There are different flavours in which cloud computing can be utilized. Depending upon the user requirements, they can exploit any of these forms of Cloud computing. The different forms of cloud computing are:

Public Cloud: In public cloud different computing resources such as servers, database and network can be accessed publicly and is available for usage to the general public. A user can either be charged for using a public cloud or it can at times be accessed without any charges. Amazon web services and Google AppEngine are some of the public clouds that are available for usage.

Private Cloud: Private cloud is a cloud which has been set up by a particular organization for its own private usage. Therefore, customized private cloud setups can be fostered for varied organizations depending upon their needs and applications. The users of a private cloud generally include people from within an organization. The level of usage amongst the different employs might also be restricted generally depending upon the relevance of the employee to the project. Organizations such as Google and Microsoft have successfully established their own private cloud [14].

Community Cloud: Community cloud is a cloud setup where a cloud can be used by multiple organizations that might have common interests in terms of security, resource requirements and common software concerns. This arrangement provides several monetary benefits to the organizations by sharing the initial cloud setup costs. Google Gov is a community cloud setup which has been done by Google.

Hybrid Cloud: Hybrid cloud is a cloud which can be a combined form of two or more kinds of cloud. It can be a mixture of public or private and community or private. For example an organization might use a blend of Amazon S3 along with some of its own in house cloud. Apart from the different flavours of cloud computing. Cloud can be offered as service in form of different service delivery models such as software as a service, plat-

- *Mansaf Alam is with Department of Computer Science, Jamia Millia Islamia (A Central University), New Delhi, India. E-mail: malam2@jmi.ac.in.*
- *Kashish Ara Shakil is with Department of Computer Science, Jamia Millia Islamia (A Central University), New Delhi, India. E-mail: shakilkashish@yahoo.co.in..*

***Please provide a complete mailing address for each author, as this is the address the 10 complimentary reprints of your paper will be sent***

*Please note that all acknowledgments should be placed at the end of the paper, before the bibliography (note that corresponding authorship is not noted in affiliation box, but in acknowledgment section).*

form as a service, database as a service and infrastructure as a service. Now this concept has been extended further towards implementing everything as a service (XaaS), where all the computing resources can be provided as a service to the cloud users. This paper concentrates on IaaS level and proper allocation and monitoring of the resources at this level.

Virtualization techniques acts as the backbone of cloud technology, involve constant allocation and reallocation of resources. This allocation and reallocation of services is done at the service provider level [2] and provides the clients with a view of an infinite compute and storage capacity. The service providers are constantly flooded with demands for high performance computing and newer resources. This constant and varying demand for resources by the clients is what lead to the development of novel scheduling techniques across a distributed system such as cloud. These scheduling algorithms have to also take into account the fluctuating demands of the clients and availability of resources at the service provider level.

Many companies have come up with a varied range of cloud services over the internet, therefore now the choice of accurate, efficient and suitable service based on the consumer requirements has become a challenging problem for the decision makers [7]. SMI Cloud [7] provides a framework for ranking and comparing different cloud services to the cloud users. Apart from these there are other approaches such as multi criteria decision making (MCDM) [8] and techniques like AHP (analytic hierarchy process) which have been explored for ranking and selection of different cloud services. These initiatives are meant to aid in decision making for the cloud users but no concrete efforts have been made for decision making at the service providers side. Decision making at the service providers side includes allocation of resource, distribution of workload, catering to the dynamic resource requirements by the consumers and consumer prioritization. All these decision making problems especially allocation of resources is very important for the service providers because efficient allocation of resources is the most fatal aspect in order to provide cloud services. Therefore the given framework is an attempt towards helping service providers for making important decisions such as allocation of resources for the service providers with the help of a decision making matrix which further aids the service providers in making decisions regarding allocation of resources using a priority based consumption value. An NDBMMMM (Network Bandwidth Dependent DMMM) algorithm has also been introduced which is based on the already existing DMMM [5] (decision matrix based max-min algorithm) algorithm. This algorithm assigns the resource having maximum value to the task which takes minimum total time. In order to allocate the resources in an optimal manner and to maximize profits it is necessary to properly analyze the available resources at disposal and then distribute and redistribute them. In order to achieve this, in this paper analysis of Google cluster trace has been done, which can be used for simulating resource usage and workload pattern in a cloud environment.

The remainder of this paper is organized as follows. Section 2 gives an insight about work already done in literature relating to the proposed framework. In section 3 models that have been used in this paper are defined, followed by section 4 which describes the proposed framework. Section 5 shows how the implementations of the proposed framework were carried out. In section 6 performance evaluation of NBDMMMM algorithm is done based on number of tasks and task arrival interval. Section 7 discusses the advantages and benefits of the framework. Finally the paper concludes with the conclusion and future works in section 8.

## 2 RELATED WORK

Scheduling of tasks, proper assignment and mapping of resources is an integral part of cloud computing. Monitoring the activities and performance of cloud is equally important in order to ensure proper resource provisioning. This monitoring of activities of cloud involves twin perspective i.e. Cloud service providers perspective and cloud users perspective [9].

Cloud service providers monitor activities like allocation of resources and meeting the end users demand. The end users on the other hand monitor the quality of service which is being provided to them apart from data security and safeguarding data against potential threats. There are two categories under which performance monitoring can be classified as per Vineetha V. [11], Infrastructure performance and Application performance. Infrastructure performance involves measuring the performance of cloud resources that are provided as a service to the cloud users, these may include network, storage and servers. Application performance management involves monitoring of databases and applications that provide support for application program performances. According to G. Aceto et al. [10] there are several activities for which resource monitoring are important such as planning the resources and cloud capacity, management of resources and data centers, security and performance management. Benefits associated with performance monitoring are also significant. Some of the benefits are identifying the adequate number of resources required for supporting a particular client, identifying lack of resources, better value for money, identifying potential performance issues [12] and choice over variety of services being offered.

Resource management and allocation is possible in cloud with the help of technology of virtualization. Virtualization offers its users complete transparency but this transparency has now resulted in further complications in terms of distribution of resources and flexibility. The characteristics for task scheduling and resource allocation in cloud as enlisted by H.Sun et al. [13] are :a) cater to distribution of resources of a unified platform. This platform may involve all the different types of PC's, workstations and servers b)task scheduling in cloud must be globally centralized c) independent scheduling of every node in cloud c) task scheduling must cope up to the scalability feature of cloud computing 4) support for dynamic scheduling depending upon the increase or de-crease in demand for the number of resources 5) scheduling strategies must proceed in sets which involves scheduling of cloud applications and scheduling of port resources.

IS. Moreno et al. [17] have done an analysis, modeling and simulation of workload patterns in utility clouds of very large scale. They have carried out their study using cloud data centers having approximately 900 users who submit 25 million tasks in one month. They have modeled by extending the capabilities of Cloud Sim framework. Their work provides a platform for researchers to simulate resource consumption patterns in the production environments. They also made several conclusions such as workload is dependency of workload on user behavior along with tasks, higher degree of diversity exists in user's pattern as compared to task patterns.

C. Tsai et al. [20] have proposed a Hyper-Heuristic scheduling algorithm for providing scheduling solutions in a cloud based environment. They have implemented their algorithm on CloudSim and Hadoop.

K. Konstanteli et al. [19] have come up with a probabilistic model for optimizing allocation of resources. The output of this model is in the form of allocation pattern which includes hosts, their subnets, storage and computational capacity. They have also used affinity and anti affinity rules. A heuristic solver has also been used for cutting down computation times and is used at runtime to scale the resources up and down.

M.A. Rodriguez and R. Buyya in [18] have put forward a particle swarm optimization based algorithm which meets the deadline constraint and minimizes the execution cost of workflows. They have represented task in a workflow as particle and number of tasks as its dimensions. The range of movement of coordinates depends upon the number of resource. They have evaluated their approach using CloudSim.

# 3 PROBLEM INCEPTION

In this section, models and parameters that have been used in this paper are defined. First, NBDMMM scheduling model is described followed by Task model and then parameter used has been specified.

## 3.1 NBDMMMM Scheduling Model

NBDMMMM scheduling model assumes that there are infinite amount of machines that are available at the service providers side. These machines are provisioned to the clients as soon as a request is made depending upon the user types and their respective priorities.
The machines available can be defined as a set M of machines where $M_i$ = {$M_1$, $M_2$...} where 'i' can be infinite.
Each of these machines $M_i$ is characterized by different resource types such as memory (m), bandwidth (b) and CPU (c).Thus, each $M_i$ = f (m, b, c). Every service provider contains several machines in form of virtual machines with varying resource types. Now these machines at the service provider side are assigned to the clients using algorithm 2, i.e. NBDMMM algorithm.

## 3.2 Task Model

Tasks arriving at the service provider's level can be defined as a set T= {$t_1$, $t_2$...}. Each of these tasks submitted by a user can be defined in terms of parameters i.e. $t_i$ = {$n_i$, $e_i$}, where $n_i$ and $e_i$ are network latency and execution time respectively. Network latency will be described further in section 5.

In order to find out execution time we assume that the tasks arriving at the service provider's side are of varying lengths with different resource usage requirements and of different durations. Let $l_i$ denote duration of each task and $c_i$ denote amount of CPU usage by a particular task. Thus, $e_i$ = $l_i / c_i$                                                          (1)
From network latency and execution time we can calculate the total time taken by a task thus,

$$Total\ time\ = n_i + e_i \qquad (2)$$

### Parameters

| | |
|---|---|
| $M_i$ | machines available |
| m | memory |
| b | bandwidth |
| c | cpu |
| T | total tasks |
| t | available tasks |
| $n_i$ | network latency for each task |
| $e_i$ | execution time for each task |
| $l_i$ | duration of each task |
| $x_a$ | x-coordinate position of cloud users |
| $y_a$ | y- coordinate position of cloud users |
| $x_i$ | x-coordinate position of the data center |
| $y_i$ | y-coordinate position of the data center |
| $d_i$ | distance between cloud users and datacenter |
| r | resources |
| v | value associated with each resource |
| RR | resource utilization rate |
| $|H|$ | Number of hosts |
| $|T|$ | Number of Tasks |

# 4 PROPOSED FRAMEWORK

Cloud computing has now gained interest from not only information technology experts but also from people belonging to other fields and domains such as scientists, engineers and doctors. Many of the applications of cloud depend upon Infrastructure as a service. Management of resources that are provided in a cloud environment is very important as the amount of resources that are required is not fixed and depends upon the cloud users, if these resources are not managed properly then any organization providing cloud services can face a major setback and will not be able to provision resources to all of its potential clients. Therefore, there is a need for constant monitoring of cloud infrastructure and improving its performance which can be achieved if the resources are distributed in an efficient manner. Furthermore, in order to manage these resources properly there is also a need that resource allocations and other decisions that are taken by the service providers should have explored all the aspects from both the service provider's side as well as the consumer's side. The NBDMMM framework helps the service providers in making decisions regarding allocation of resources through NBDMMM algorithm and a decision matrix. The proposed framework is based on DMMM algorithm according to which improvement in infrastructure performance can be made using a decision matrix based max min algorithm (DMMM). In this algorithm monitoring techniques are used along with DMMM algorithm. The proposed framework involves cloud service providers and their monitoring techniques, cloud users and an improved network bandwidth dependent DMMM algorithm (NBDMMM). The advantage of the proposed algorithm over the former DMMM algorithm [5] is that it also takes into account network bandwidth during allocation of resources. Fig. 1 shows a broad outline of the proposed approach. The various actors in the given framework therefore include cloud service providers and client or end users using cloud services. The major key to this algorithm is to achieve an efficient, cost effective and robust technique for allocation of resources that provides its users with high availability of resources such that all the resource requirements of cloud users are met at all times. Resource allocation in cloud computing is a non cooperative problem as the consumers who wish to access the same resources are competitors and thus reluctant to cooperate with each other [1].Therefore, there is a need to make sure that demands of all the consumers are met at all times and no consumer is left starving for resource. This can only be achieved with an efficient resource allocation strategy that caters to the needs of all its clients. The usage of cloud resources by a consumer is not fixed and varies depending on the consumer requirements. This variation in usage pattern is monitored by the framework and depending upon this pattern and various other parameters decisions regarding assignment of these resources is done by the service providers. The key elements of the framework therefore include the following:

1. Service Brokers: This component has the responsibility of coordinating with cloud users, interacting with them and understanding their requirements and needs. It also keeps track of the fact that whether a customer has certain special privileges or requirements which is helpful in prioritizing these customers.

2. Monitor: This component is responsible for monitoring the activities of end users along with usage pattern of resources being consumed. It takes into account the frequency of usage of machines, number of VMs required, bandwidth usage, scalability requirements and hours of peak as well as dormant demands.

3. Resource Allocator: This component is responsible for allocating resources to the cloud users depending upon the resource availability and resource usage pattern. NBDMMM algorithm is applied at this component level.

Thus the important concerns in building of this framework involve, assessing the usage pattern and assignment of priorities to the cloud users. Infrastructure performance enhancement using the proposed framework

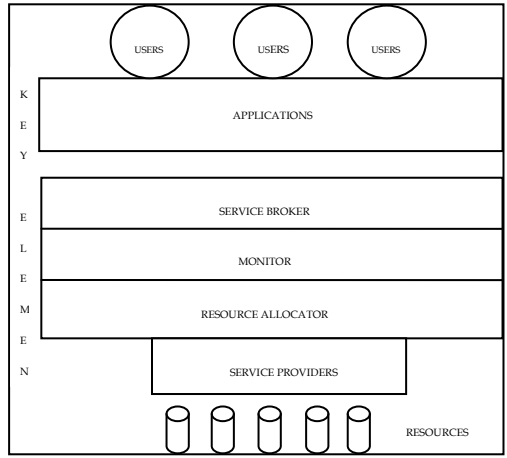

Fig. 1. Proposed Framework

follows the following stages:

Stage 1: Cloud service provider monitors the infrastructure level access pattern of resources and behavior of the cloud users. This task is carried out by the monitor module.

Stage 2: Report the monitoring activities about the usage to cloud service providers.

Stage 3: Apply proposed Network Bandwidth Dependent DMMM algorithm (based on DMMM algorithm). The information collected by service brokers is used at this stage.

Stage 4: Allocate resources or provide services to cloud users, thereby leading to infrastructure level performance enhancement and efficient management of resources.

## 4.1 Stage 1: Cloud service provider monitors the infrastructure level access pattern of resources and behavior of the cloud users

Cloud users tend to use resources or services in a dynamic manner depending upon their requirements. Sometimes the usage of resources is more and reaches its peak value and sometimes it's less or almost negligible .Therefore, at this stage monitoring of the usage pattern of the way in which these resources are exploited by the consumers is done. These resources can be servers, virtual machines, storage and software bundles [2].This task is carried out by the monitor module. This monitoring of activities helps the service providers determine the access pattern of resources, trends by which data and resource access is made and also hierarchy of access of data which in turn can help in generating conclusive information's such as the hours of peak demands or least demands against resources by the cloud users. It acts as an indicator for the service providers to adjust their allocation of resources to its clients depending on their usage pattern. For example the graph given below in Fig. 2 provides the information about the usage pattern of resources by a particular cloud user.This usage pattern can then be used for developing a monitoring report. According to this monitoring report, the maximum resource requirement or peak demand by this client is encountered at 09:00 hours and at 10:00 hours this requirement is the least or negligible i.e. no resources are required at this time. So, now the cloud service providers have to adjust their distribution of resources according to the number of resources demanded by each of its clients. This step is one of the most crucial step because if there are any faults in the monitoring activities and if the access pattern of resources has not been assessed properly then guaranteeing optimal distribution of resources by the service providers will fail.

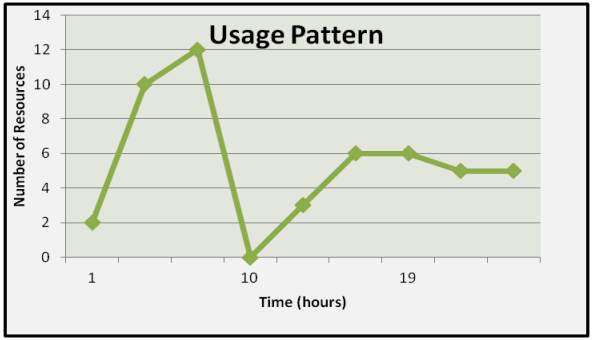

Fig. 2. Usage Pattern of resources by a Particular Cloud User

### 4.2 Stage 2: Report the monitoring activities about the usage to cloud service providers

This stage involves reporting about the access pattern of resources to the service providers. After the information regarding the usage pattern has been gathered or organized as a report. This, report is then sent to the cloud service providers. The team at the cloud service provider's site which includes the cloud database administrator and data center programmers [3] then decide upon how the available resources are to be distributed amongst the multiple clients. This decision is based upon the proposed NBDMMM algorithm and consumption values or user priority. It is at this juncture when the service provider can detect the time of dormant activities by its client and thereby attract more clients perhaps from a different time zone. This leads to a mutual beneficial situation for both the service providers as well as the clients. The service providers benefit by utilizing their resources in an optimum manner and the clients benefit by getting better service availability.

### 4.3 Stage 3: Apply proposed Network Bandwidth Dependent DMMM algorithm (based on DMMM algorithm)

Cloud end users require resources in an elastic manner depending on their needs, due to this flexibility in the requirement of the number of resources, task scheduling mechanisms of grid computing cannot work efficiently in cloud based environments [4]. Therefore, we need newer algorithms for distribution of resources in cloud which are developed specially for cloud based environments. Our proposed NBDMMM Algorithm allocates resources to tasks in a cloud environment. This optimal allocation of resources eventually leads to infrastructure performance enhancement for the service provider who is providing the cloud environment to its users. It also has several performance benefits and outperforms many existing algorithms for resource allocations.

The proposed NBDMMM (network bandwidth dependent DMMM algorithm) is based on the already proposed DMMM algorithm [5]. DMMM algorithm was based on min-min scheduling algorithm. According to DMMM algorithm shown in Algorithm 1, if T = {$t_1$, $t_2$..........$t_n$} be 'n'cloud users tasks, that are to be assigned resources R = {$r_1$, $r_2$..........$r_m$}, where 'm' is the number of resources .If Xij is a value calculated from {$v_1$, $v_2$.........$v_k$} as the outputs of a decision matrix where Xij = maximum value of {$v_1$, $v_2$.......$v_k$}. Then DMMM algorithm selects the resource with value of $X_{ij}$ = max ($v_1$, $v_2$.......$v_k$) and assigns this resource to task which takes minimum time for its execution.

---

ALGORITHM 1
DMMM ALGORITHM

---

1. **For** all Tasks $t_i \in T$
2.    **For** all resources $r_j \in R$
3.    $X_{ij}$ =max($v_1v_2$......$v_k$)
4.    **End** For;
5. **End** For;
6. **Do** while T is not empty
7. Find task $t_i$ with minimum execution time
8. Assign task $t_i$ to resource $r_j$ having $X_{ij}$ as value
9. Remove task $t_i$ from T
10. **End** Do

---

The proposed NBDMMM algorithm is shown in algorithm 2 is based on DMMM algorithm. NBDMMM algorithm first finds the resource having maximum value, it then finds out the task that requires minimum total time. Here minimum total time is a function of network bandwidth and execution time. This algorithm will iterate till all the tasks have been assigned resources.

---

ALGORITHM 2
NBDMMM ALGORITHM

---

1. **For** all resources $rj \in R$
2. $X_{ij}$ =max($v_1$, $v_2$.......$v_k$)
3. **End** For;
4. Sort $X_{ij}$ in ascending order
5. **Do** while T is not empty
6. Find task $t_i$ with minimum total time ,
Where, total time = network latency+ execution time
7. Assign task $t_i$ to resource $rj$ having $X_{ij}$ as value
8. Remove task $ti$ from T
9. **End** Do

---

Fig. 3 shows placement of cloud users and data centers at varied geographical locations. There can be several cloud users querying different data centers with their requirements. These location co-ordinates have been explored further in our study so as to calculate total time in execution of a request by the client in the proposed framework.

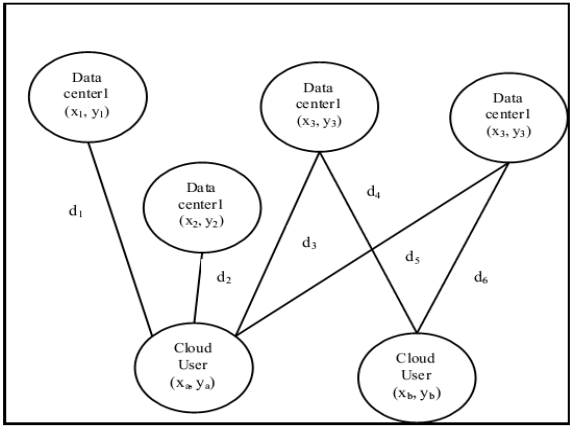

Fig. 3. Placement of Cloud Users and data centers

Here total time is a function of network latency, execution time and time taken for switching the server on and off, but in this algorithm server switch time has been ignored. Here,

Total time = network latency + execution time        (1)

Total time is defined as the sum of execution time and network latency. Where, execution time is the time taken by a user to complete its operations or tasks and network latency is described in terms of time and distance.

Let $(x_a, y_a)$ be coordinate positions of cloud users and $(x_i, y_i)$ be coordinates of the data center, both may be dispersed across several geographical locations.

Let $d_1, d_2...d_n$ be distance of data center$_1$, datacenter$_2$ and so on be distance of data centers from cloud user.

Therefore distance $d_i$, where 0<i<n+1 is given and 'n' is the total number of data centers owned by cloud service providers dispersed at varied geographical locations is given by (3).

$$d_i = \sqrt{((x_i - x_a)^2 + (y_i - y_a)^2)}$$        (4)

Now distance $D = \min \{d_1, d_2, d_3 .... d_i\}$        (5)

Thus, Network latency is given by (5):

Network latency=(distance D)/(network bandwidth)    (6)

Where, Network bandwidth is defined as speed of network in bits per time units.

For this proposed algorithm, decision matrix has been used as a tool for making decisions regarding distribution of resources depending upon the user priorities and consumption value of resources along with total amount of resources held at disposal of the service providers. Here the various users have been classified as casual users, underprivileged users, naive users, privileged users and high end users. The information regarding priorities is gathered by the service broker module.

**4.4 Stage 4: Allocate resources or provide services to cloud users, thereby leading to**

**infrastructure level performance enhancement and efficient management of resources**

After cloud users have been assigned priorities and their respective values have been calculated and after application of NBDMMM algorithm, resources are then distributed to the end users depending on their rights and service provider's driven priority values. This leads to mutual benefits for both the service providers and the users thereby leading to a win –win situation for both the consumers and service providers. In the proposed framework a value is associated with each of the resource. This value is based on several factors such as costs, deadline, length of tasks and priorities. In order to meet this end a decision matrix has been used which can be used for calculating the values associated with each of the resources thereby helping in allocation of resources. This decision matrix thus helps the service providers in their decision making tasks. In the proposed framework the priority factor is further explored .The priority is thus described as a factor of consumption value. i.e.

Priority $\propto$ Consumption value        (3)

## 5 IMPLEMENTATIONS

In order to validate the proposed framework implementations were carried out using data from Google cluster [16] and cloud analyst. Cloud analyst is a cloud based simulator and has been used in our experiments for calculating network execution time.

### 5.1 Format of Data Set for Testing

We have used data set provided by Google [16]. This data set consists of traces of production workloads that were acquired after running on Google cluster for about 29 days. The Google cluster consists of machines that are connected by a very high bandwidth cluster network. The total number of machines used are about 12,000.The workload is divided into a set of jobs; each job consists of one or many tasks. Each of these tasks consists of Linux programs with multiple processes. The data set is divided into 6 tables namely Machine Events table, Machine attributes table, Job Events table, Task events table, Task constraints table and task resource usage table. Out of these tables Jobs events table and task resource usage table has been used for our work.

Step 1: Cloud service provider monitors the infrastructure level pattern of usage of resources and behavior of the cloud users.

The service providers will first monitor the usage pattern of the cloud users that is what all resources are being needed by the users and for how long, after this the service providers will have a fair idea about the usage pattern of the customers. Therefore at the end of this step the service providers will have knowledge such as execution time and hours of peak demand along with the hours during the day when the demand is least .This has been summarized in Table I. In order to simulate the resource usage pattern for our framework we have used data from Google cluster trace for this we have used Jobs event table and task resource usage table, after evaluation of the usage pattern, a summary was prepared. There were around 497 jobs each containing about 10705 tasks approximately. Table 1, shows an illustrative example of resource usage pattern for first 50 tasks for a job. This pattern collected from Google cloud trace has been used only for accessing the probable behavior of the cloud users however cloud analyst which is a cloud simulator has been used for calculating execution time for results. Tasks field in the given table represents tasks for carrying out different jobs, each job can have multiple tasks associated with it. Job ID field represents the job id of a particular job, Event types represent the particular task or job events .event type 0 means a job is eligible for scheduling, 1 means task or job is scheduled on some machine, 2 means Evict i.e. job is preempted, 3 means Fail due to task failure, While 4 represents Finishing of a job after completion and 5 represents Killing of job due to failure of a dependent job or task. The scheduling class represents latency sensitivity of tasks a higher value indicates high latency sensitivity and lower value indicates less critical task or low latency task. From the analysis of the trace data we can deduce that out of all the jobs that have been submitted for execution only half of them are actually completed, after evaluation of all the job tables the time stamp at which maximum demand for resources was made can be deduced. This pattern can later on be used for provisioning of resources.

Step 2: Report the monitoring activities about the usage to cloud service providers.

After the usage pattern has been observed and monitoring activities recorded .This information is then sent out to the cloud service providers. This information can be very vital for the service providers and will aid them in making important decisions regarding allocation and distribution of resources.

Step 3: Apply proposed Network Bandwidth Dependent DMMM algorithm

In order to apply NBDMMM algorithm the consumption values associated with each of the resources r1, r2, r3 and r4 are calculated. The values associated with each of the resources depend upon the user preferences, availability of resources at disposal, distance of the user from data centers and algorithms being used by the service providers.

TABLE 1
RESOURCE USAGE PATTERN OF GOOGLE CLUSTER

| Tasks | Time stamp | Job ID | Event type | Scheduling class |
|---|---|---|---|---|
| T1 | 2.45E+12 | 6.48E+09 | 0 | 0 |
| T2 | 2.45E+12 | 6.48E+09 | 1 | 0 |
| T3 | 2.45E+12 | 6.48E+09 | 0 | 0 |
| T4 | 2.45E+12 | 6.48E+09 | 0 | 2 |
| T5 | 2.45E+12 | 6.48E+09 | 4 | 0 |
| T6 | 2.45E+12 | 6.48E+09 | 1 | 0 |
| T7 | 2.45E+12 | 6.48E+09 | 1 | 2 |
| T8 | 2.45E+12 | 6.48E+09 | 0 | 0 |
| T9 | 2.45E+12 | 6.48E+09 | 0 | 2 |
| T10 | 2.45E+12 | 6.48E+09 | 4 | 2 |
| T11 | 2.45E+12 | 6.48E+09 | 5 | 0 |
| T12 | 2.45E+12 | 6.48E+09 | 5 | 2 |
| T13 | 2.45E+12 | 6.48E+09 | 1 | 0 |
| T14 | 2.45E+12 | 6.48E+09 | 1 | 2 |
| T15 | 2.45E+12 | 6.48E+09 | 4 | 1 |
| T16 | 2.45E+12 | 6.48E+09 | 0 | 1 |
| T17 | 2.45E+12 | 6.48E+09 | 4 | 0 |
| T18 | 2.45E+12 | 6.48E+09 | 1 | 1 |
| T19 | 2.45E+12 | 6.48E+09 | 0 | 1 |
| T20 | 2.45E+12 | 6.48E+09 | 0 | 1 |
| T21 | 2.45E+12 | 6.48E+09 | 0 | 1 |
| T22 | 2.45E+12 | 6.48E+09 | 1 | 1 |
| T23 | 2.45E+12 | 6.48E+09 | 1 | 1 |
| T24 | 2.45E+12 | 6.48E+09 | 1 | 1 |
| T25 | 2.45E+12 | 6.48E+09 | 4 | 2 |
| T26 | 2.45E+12 | 6.48E+09 | 5 | 0 |
| T27 | 2.45E+12 | 6.48E+09 | 5 | 2 |
| T28 | 2.45E+12 | 6.48E+09 | 4 | 1 |
| T29 | 2.45E+12 | 6.48E+09 | 0 | 0 |
| T30 | 2.45E+12 | 6.48E+09 | 1 | 0 |
| T31 | 2.45E+12 | 6.48E+09 | 0 | 2 |
| T32 | 2.45E+12 | 6.48E+09 | 0 | 1 |
| T33 | 2.45E+12 | 6.48E+09 | 0 | 1 |
| T34 | 2.45E+12 | 6.48E+09 | 0 | 0 |
| T35 | 2.45E+12 | 6.48E+09 | 1 | 2 |
| T36 | 2.45E+12 | 6.48E+09 | 0 | 0 |
| T37 | 2.45E+12 | 6.48E+09 | 0 | 2 |
| T38 | 2.45E+12 | 6.48E+09 | 1 | 1 |
| T39 | 2.45E+12 | 6.48E+09 | 1 | 1 |
| T40 | 2.45E+12 | 6.48E+09 | 1 | 0 |
| T41 | 2.45E+12 | 6.48E+09 | 1 | 0 |
| T42 | 2.45E+12 | 6.48E+09 | 1 | 2 |
| T43 | 2.45E+12 | 6.48E+09 | 4 | 0 |
| T44 | 2.45E+12 | 6.48E+09 | 3 | 1 |
| T45 | 2.45E+12 | 6.48E+09 | 4 | 0 |
| T46 | 2.45E+12 | 6.48E+09 | 0 | 0 |
| T47 | 2.45E+12 | 6.48E+09 | 1 | 0 |
| T48 | 2.45E+12 | 6.48E+09 | 4 | 1 |
| T49 | 2.45E+12 | 6.48E+09 | 4 | 0 |
| T50 | 2.45E+12 | 6.48E+09 | 4 | 0 |

This information is usually deduced from the usage pattern by the experts at the service provider's site. Priority plays a major role in calculation of this value i.e.Priority ∝ consumption value . as per (1).For illustration purpose five user types have been defined namely casual users, underprivileged users, naive users, privileged users and high end users. The priority value five is the highest priority value and 1 is the lowest priority value. The user types and their respective priorities have been elicited in Table 2. As per the criteria's, constraints and preferences of the service providers a decision matrix will be drawn given by Fig. 4. The usage of decision matrix as a tool for calculating the value for each of the resources is that it allows the service providers to assign values keeping in mind their different criteria's and constraints. Similarly value corresponding to other resources can be found out

| USER TYPES | | | | | | |
|---|---|---|---|---|---|---|
| | | HIGH END USERS | PRIVILEGED | CASUAL USERS | NAÏVE USERS | UNDER PRIVILEGED |
| CRITERIA'S | CONSTRAINTS / WEIGHTS | PRIORITY=5 | PRIORITY=4 | PRIORITY=3 | PRIORITY=2 | PRIORITY=1 |
| CRITERIA C1 | 1 | 1*5 | 1*4 | 1*3 | 1*2 | 1*1 |
| CRITERIA C2 | 2 | 2*5 | 2*4 | 2*3 | 2*2 | 2*1 |
| CRITERIA C3 | 3 | 3*5 | 3*4 | 3*3 | 3*2 | 3*1 |
| TOTAL | $V_i$ | 30 | 24 | 18 | 12 | 6 |

Fig. 4. Calculating value for a particular resource 'A' using a decision matrix

### TABLE 2
#### User Types And Their Respective Priorities

| USER TYPES | PRIORITY |
|---|---|
| High end users | 5 |
| Privileged users | 4 |
| casual users | 3 |
| Naïve users | 2 |
| under privileged users | 1 |

### TABLE 3
#### Network Latency, Execution Time and Total Time

| Tasks | Network Latency(ms) | Execution time(ms) | Total Time(ms) |
|---|---|---|---|
| T1 | 50 | 0.88 | 50.88 |
| T2 | 10 | 0.54 | 10.54 |
| T3 | 15 | 0.33 | 15.33 |
| T4 | 30 | 0.90 | 30.90 |
| T5 | 25 | 0.64 | 25.64 |

easily using the decision matrix method. Let these values be 30, 40, 95 and 105 respectively for resource a, resource B, resource C and resource D respectively. Now, on the basis of the decision matrix the maximum value of re sources is found from amongst the value of resources i.e.
$X_{ij}$ = max {30, 40, 95, 105} = 105
After the maximum values associated with each of the resources is found out we calculate the total time required by each of the tasks

*Total Time = Network latency + Execution time* (2)
If the network latency for each of the tasks is given as in Table 3. Total time can be calculated from execution time and network latency using (2), (4), (5) and (6).
In order to calculate execution time based on the distance from the data centre's cloud analyst [15] was used which is a cloud simulator for carrying out simulations and knowing the distribution of resources across different data centers. Simulations were carried out for each of the tasks for example to calculate execution time for task T1, Simulations were done using cloud analyst simulator with three data centers DC1, DC2 and DC3 and one cloud user UB1. Closest data centre was taken as the service broker policy, as the requirement was for data centre having minimum distance from the cloud user. Fig. 5 given below shows the initial location of each of the data centers and users across dispersed geographical location.
Similarly for tasks T2, T3, T4 and T5 simulations were performed for calculating execution time. For the illustrative example maximum response time has been taken as the execution time. The network latency values were fed into the transmission delay between regions matrix of the cloud analyst. The main configurations for the simulation have been given by Fig. 6. Since data centre D3 is closest to the user therefore it's chosen and execution time is calculated at D3.

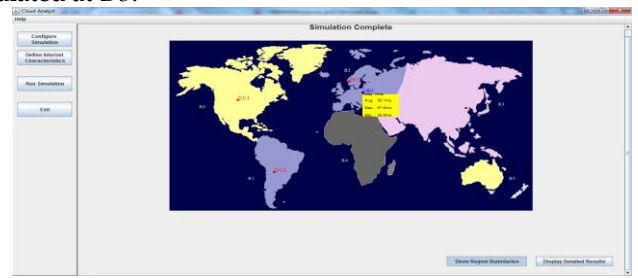

Fig. 5. Initial location of Cloud users and data centers

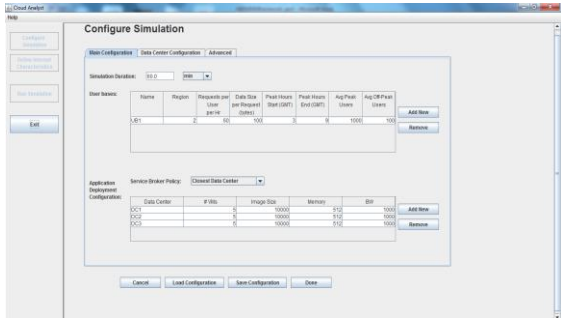

Fig. 6. Main configuration settings for simulations using cloud analyst

Data Center Request Servicing Times

| Data Center | Avg (ms) | Min (ms) | Max (ms) |
|---|---|---|---|
| DC1 | 0.00 | 0.00 | 0.00 |
| DC2 | 0.00 | 0.00 | 0.00 |
| DC3 | 0.50 | 0.02 | 0.88 |

Fig. 7. Total time for tasks

Now from the given Total time for each of the tasks minimum total time task is found out in Fig. 7 , Here task T2 takes the least time and is assigned the resource with maximum value i.e. resource with value 105.

## 6 PERFORMANCE EVALUATION AND RESULTS

In order to exhibit the performance efficiency obtained by NBDMMM algorithm we compare it with three existing algorithms Greedy-R [21], Greedy-P [21] and FCFS [21].
Greedy-R (Greedy Response) scheduling: In order to maximize the response time of a system the task having quickest time of execution first is assigned to the cloud resource which is most powerful.
Greedy-P (Greedy Parallelization) scheduling: In order to maximize the response time of a system and achieve task parallelization the task having quickest time of execution first is assigned to the cloud resource which is least powerful.
FCFS (First come first serve) scheduling: In this scheduling tasks are assigned to any of the available cloud resources as soon as they arrive.
The experiments have been conducted using CloudSim toolkit [22] which is a simulation framework for cloud computing environments. As compared to other simulation toolkits such as SimGrid and GangSim it helps in modeling of virtualized resources on demand [23].
We have simulated the experiment on 20 nodes. Each node has one CPU core with performance of 2000, 2500 and 3000 MIPS, 4 GB RAM and 100 GB storage. Each host can have multiple VMs with one CPU core, 64 MB RAM and 256 MB storage.
The performance has been calculated by using the following metrics. These metrics are similar to the ones used by Z. Xiaomin et al. in [24]:

- Task Success Ratio: Task success ratio is defined as ratio of tasks executed/ total number of Tasks submitted.
- Resource Utilization Rate: The resource utilization rate at each host node can be calculated as

$$\text{RR} = \left( \frac{\sum_{k=1}^{|H|}(\sum_{i=1}^{|T|} l_{ik})}{\sum_{k=1}^{|H|} c_k} \right) \quad (7)$$

Where $|H|$ is the number of hosts
- Task Arrival Interval: Task Arrival interval determines the difference in arrival time between two tasks.

### 6.1 Performance Impact of Number of Tasks
A series of experiments were conducted in order to find out the impact of the difference in number of tasks on the performance of NBDMMM algorithm. We also compare the performance of NBDMMM algorithm with FCFS, Greedy-R and Greedy-P algorithm. From Fig.8 (a) we infer that Task success ratio of all the algorithms remain same despite the change in the number of tasks. This uniformity in the success ratio can be attributed to the scalable and elastic nature of cloud environment. As the number of tasks being submitted by the cloud users increases the number of resources being made available to the users also increase. Fig.8 (a) also shows that NBDMMM algorithm has higher task success ratio in comparison to other algorithms. From Fig.8 (b) it can be observed that NBDMMM algorithm provides better utilization of resources in contrast to the other four algorithms. Percentage of resources being consumed by NBDMMM algorithm is higher than others. This better utilization of resources can be due to taking into consideration network latency during assignment of resources to different tasks.
This better utilization of resources can be due to taking into consideration network latency during assignment of resources to different tasks. Thus, NBDMMM algorithm proves to be a more optimal solution for resource allocation as it helps in allocation of resource in an efficient and optimal manner. Further analysis of Fig. 8 (b) is shown in Table 4, from this it be deduced that NBDMMM shows less percentage error as compared to the other three algorithms i.e. 6.68 percent while percentage error in Greedy P, Greedy R and FCFS is 23.25, 23.38 and 17.67 percent respectively. The covariance value in case of NBDMMM algorithm is also high  and resource consumption is positively correlated to the number of tasks. We can also observe that there exists a stronger correlation (adjusted $R^2$ = 0.928) between number of tasks and resource consumption in case of NBDMMM algorithm as compared to others.

### 6.2 Performance Impact of Task Arrival Interval
The rate at which tasks arrives at the service provider also has an impact on the performance. Therefore to assess the affect of task arrival interval we have taken the value of tasks between the range of [0, 20]. Fig. 9 reveals the affect of arrival interval of tasks on its success ratio and resource consumption.
From Fig. 9 (a) it can be observed that the task arrival interval do not have a significant impact on the task suc cess rate. The task success rate is almost similar to the ones in Fig.8 (a) this is due to the availability of large number of resources at the service provider's end which are made available to the consumers dynamically.

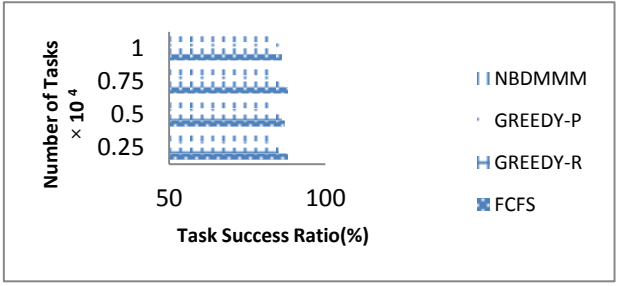

(a)

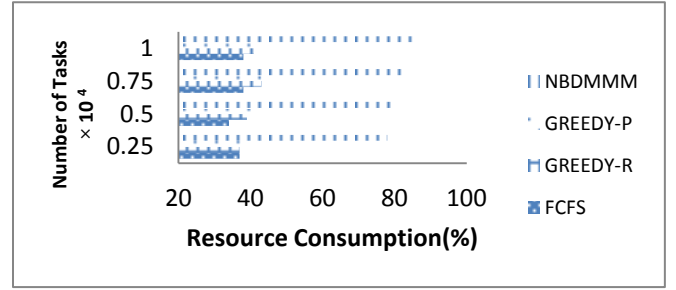

(b)

Fig. 8. Performance Impact of Number of Tasks

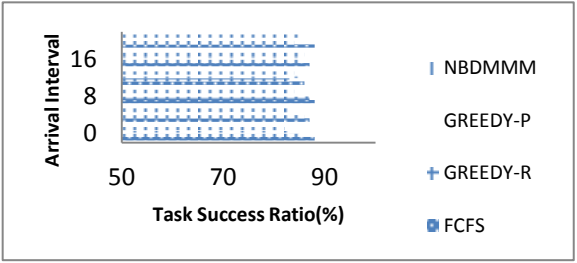

(a)

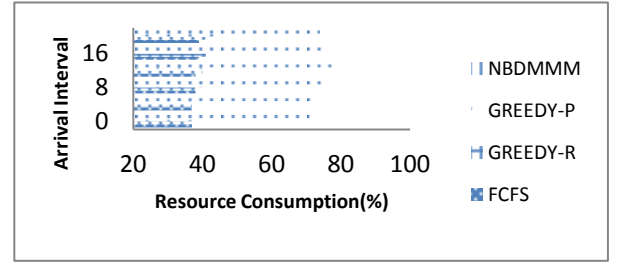

(b)

Fig. 9. Performance Impact of task Arrival Interval

From Fig. 9 (b) it can be demonstrated that NBDMMM algorithm still performs better than the other three algorithms when the task interval rate is changing. Though the percentage by which it outperforms others is a little less than the one in Fig.8 (b)

## 7 ADVANTAGES AND BENEFITS OF THE PROPOSED FRAMEWORK

The proposed framework aims at providing several benefits to both the service providers as well the cloud users Existing task scheduling algorithms of grid computing fail to assign resources in a dynamic environment such as cloud where the resources are assigned in an elastic manner. The proposed framework is designed especially for a cloud environment and takes into account the elastic nature of cloud.

- Use of decision matrix as tools helps in calculating the optimal value of resources depending upon the different criteria and factors.
- Even though using decision matrix as tool can be a manual approach. This tool can be easily modeled into an automatic version.
- Use of decision matrix provides plenty of scope for service providers to modify the behavior of allocation and reallocation of resources.
- NBDMMM algorithm takes into account network latency which provides the actual time required for execution of tasks and thus proves to be more lucrative and realistic approach.
- It has better performance than other allocation

algorithms like Greedy P, Greedy R, DMMM and FCFS as it also takes into account network bandwidth and other network overheads involved in transfer of data from data centers to the various end users.

Some of the benefits of the proposed framework are:

- Performance enhancement at infrastructure level: The cloud service providers distribute resources amongst the cloud users in an elastic manner after careful evaluation of the usage pattern. This leads to infrastructure level performance enhancement.
- Win–Win Situation: The proposed framework serves as a mutual beneficial platform for both the service providers as well the cloud users. Service providers benefit by better allocation of resources and better business opportunities on the other hand cloud users also benefit in terms of improved services.
- Optimal Allocation of resources: The proposed method guarantees to be an optimal resource distribution approach and helps in better allocation of resources to the cloud users.

TABLE 4
Analysis of Resource Consumption v/s Number of Tasks

| Regression Statistics | Greedy P | Greedy-R | FCFS | NBDMMM |
|---|---|---|---|---|
| Adjusted R Square | 0.4403 | 0.4396 | 0.5 | 0.9285 |
| Covariance | 0.4062 | 0.5 | 0.2187 | 0.8125 |
| Standard Error | 0.2325 | 0.2338 | 0.1767 | 0.0668 |

- Improved Quality of Services: The proposed framework helps in distribution in a more manageable and efficient manner. Thus, providing improved quality of services.
- Better Management of Resources: The cloud resources can now be distributed amongst the end-users by the service providers depending upon their policies, constraints and criteria's thereby providing better management options for the service providers.
- Enhanced end user experience: Since, the services are provided by the services providers on the basis of a decision matrix. Therefore clients can negotiate to a certain extent the quality of service they expect for example by upgrading themselves from casual to high end users.
- More Potential customers: Monitoring the resource usage pattern of the cloud users gives the service providers an idea about hours of peak usage of resources and also time during the day when the usage of resources is minimum. This gives the service providers an opportunity to attract more potential customers during the hours of low utilization of resources.

## 8 CONCLUSIONS AND FUTURE WORKS

The proposed framework uses monitoring techniques to monitor the cloud user's resource usage pattern. It then calls NBDMMM algorithm which is an improved version of DMMM algorithm and takes into account network latency thereby being a more realistic and practical approach for distribution of resources. The proposed framework uses decision matrix as a tool for calculating the value associated with each of the resources. The usage of this decision matrix as a tool for calculating value of resources, thereby leading to distribution of resources, is a lucrative approach as it gives the service providers ample of freedom for distribution of their resources depending upon their criteria's and constraints laid by the various policies of their organization. The proposed framework has several benefits and advantages such as better management of resources, improved quality of services, infrastructure level performance enhancement and ability to attract more customers. This paper has been developed focusing infrastructure level performance improvement in mind and thus, provides a better solution for distribution of resources in a cloud based environment. This concept can also be applied to other levels such as application and client levels. The proposed algorithm is based on DMMM algorithm which in turn has been adopted form Min-Min algorithm but this proposed approach can be extended to other scheduling algorithms of similar nature such as max-min, genetic and round robin algorithm. In future we plan to implement this paper by taking into consideration energy efficiency of resources.

## ACKNOWLEDGMENT

The authors wish to thank anonymous reviewers for their suggestions and comments. It is also acknowledged that Kashish Ara Shakil is the corresponding author for this paper.

**Mansaf Alam** received his doctoral degree in computer Science from Jamia Millia Islamia, New Delhi in the year 2009. He is currently working as an Assistant. Professor in the Department of Computer Science, Jamia Millia Islamia. He is also the Editor-in-Chief, Journal of Applied Information Science. He is in editorial Board of some reputed International Journals in Computer Sciences and has published about 24 research papers. He also has a book entitled as "Concepts of Multimedia, Book" to his credit. His areas of research include Cloud database management system (CDBMS), Object Oriented Database System (OODBMS), Genetic Programming, Bioinformatics, Image Processing, Information Retrieval and Data Mining.

**Kashish Ara Shakil** has received her Bachelor's degree in Computer Science from Delhi University in 2008 and has an MCA degree (2011) as well. She is currently pursuing her doctoral studies in Computer Science from Jamia Millia Islamia (A Central University). She has written several research papers in the field of Cloud computing. Her area of interest includes database management using cloud computing, distributed and service computing.